# Quantum interference in finite-size mesoscopic rings

G. P. Papari [1,2,3*] and V. M. Fomin [4,5,6]

[1] *Dipartimento di Fisica, Università di Napoli "Federico II," via Cinthia, I-80126 Napoli, Italy*
[2] *CNR-SPIN, UOS Napoli, via Cinthia, I-80126 Napoli, Italy*
[3] *Istituto Nazionale di Fisica Nucleare (INFN), Naples Unit, via Cinthia, I-80126 Napoli, Italy*
[4] *Institute for Integrative Nanosciences, Leibniz IFW Dresden, Helmholtzstraße 20, D-01069 Dresden, Germany*
[5] *Laboratory of Physics and Engineering of Nanomaterials, Department of Theoretical Physics, Moldova State University, Str. Alexei Mateevici 60, MD-2009, Chisinau, Moldova*
[6] *Institute of Engineering Physics for Biomedicine, National Research Nuclear University MEPhI (Moscow Engineering Physics Institute), Kashirskoe shosse 31, 115409 Moscow, Russia*

*Corresponding author: gianpaolo.papari@unina.it*

Abstract:
The Ginzburg-Landau theory is used to model the order parameter of a finite-size mesoscopic ring to investigate the effects of the onset of screening currents on the transport of incoming ones. The magnetic flux breaks the symmetry of currents between input and output stubs by means of an induced spatial ordering upon diamagnetic and paramagnetic supercurrents circulating in the ring. The distribution of those screening currents drives the interference of incoming/outgoing supercurrents resulting into a sinusoidal variation of resistance as a function of the magnetic flux even if the density of quasiparticles is not modified by the external magnetic field.

1. Introduction

The transport in mesoscopic multiply connected structures, shaped as rings or cylinders, was envisaged for 70 years for harnessing the control of electrical observables through the manipulation of wave-like/quantum properties of charge carriers. These studies came up to a series of discoveries ranging from quantum metrology for a new system of units [1] to quantum computing [2] [3].
The transport in quantum regime is characterized by flowing charge carriers whose state can be described through a wave function $\Psi(\bar{r}) = \Psi_0 e^{i\varphi(\bar{r})}$ featured by a modulus $\Psi_0 = |\Psi|$ and a phase $\varphi(\bar{r}) = 1/h \int_{\gamma(\bar{r})} \bar{P}(\bar{r}') \cdot d\bar{r}'$ where $\bar{P} = \bar{p} + q\bar{A}$ is the gauge-invariant momentum [4], $h$ is the Planck constant, $\gamma(\bar{r})$ is the charge carrier path and $\bar{p}$ is the charge carrier momentum, $q$ is the charge of the carrier and $\bar{A}$ is the vector potential of the magnetic field $\bar{B} = \bar{\nabla} \times \bar{A}$ applied over the system. Charge carriers can interfere only if phase memory is preserved over a length comparable with the sample itself [5]. When charge carriers preserve phase coherence during the transport, interference processes become measurable. Reference interferometers for transport experiments are mesoscopic rings whose doubly-connected topology forces phase coherent charge carriers to gain a phase $\varphi = 2\pi \frac{\Phi}{\Phi_0} = \frac{2\pi}{\Phi_0} \oint \bar{A} \cdot d\bar{r}$ related to the flux $\Phi$ piercing the ring, $\Phi_0 = h/q$ is the magnetic flux quantum and the circulation in the integral is extended over the circumference embracing the flux area. $\Phi$ alters the phase of the wave function resulting in the modulation of the sample resistance as a function of $B$. The phenomenon of the magnetoresistance (MR) oscillations (MRO) in normal and superconducting rings are usually discussed as a manifestation of the Aharonov–Bohm (AB) quantum-interference effect [6] [7] [8] [9] [10] [11] or the Little–Parks (LP) effect explained in terms of the oscillation of the critical temperature as a function of the magnetic field [12] [13] [14].
The theoretical frames behind the MRO in normal and superconducting cases have three main common features:

(*i*) MRO are periodic functions of the magnetic flux with a period equal to the magnetic flux quantum $\Phi_0 = h/e$ or $\Phi_0 = h/2e$ for unpaired (*q*=*e*) or paired (*q*=2*e*) charge carriers, respectively. Flux periodicity is an inherent consequence of the single-valuedness of the wave function. After a round about a one-dimensional ring, the wave function must recover its value yielding [15]

$$2n\pi = \frac{1}{\hbar} \oint_\gamma (\bar{p} + q\bar{A}) \cdot d\bar{r}, \ n = 0, \pm 1, \pm 2 \ldots \quad (1)$$

(*ii*) The magnetic field changes $\bar{P}$ pushing charge carriers to circulate around the ring realizing a persistent current [16].

(*iii*) The dependence of the charge carrier momentum on the external flux $\Phi$ affects the energy spectrum of charge carriers in an infinitesimally narrow ring, which is represented as discrete energy levels [11]

$$\varepsilon_n = \frac{\hbar^2}{2m^*r^2}(\Phi/\Phi_0 - n)^2, n = 0, \pm 1, \pm 2 \ldots \quad (2)$$

where $m^*$ is the charge carrier effective mass, $r$ is the ring radius and $\Phi = B\pi r^2$.

According to eq. (2) the spectrum of a single charge carrier bounded to circulate around a circle with a fixed radius unavoidably relates the particle velocity to the amplitude of *B*. Here we propose a new approach to avoid the dependence of the charge carrier velocity on the magnetic field. The major difference with respect to the previous approaches [11] [17] arises from considering a finite-size mesoscopic ring, where the eq. (1) valid for 1D rings is approximately extrapolated to 2D rings by an appropriate evaluation of the specific path $\gamma$. Quantum interference in finite-size mesoscopic rings was rigorously analyzed using the 2D Schrödinger equation for semiconductors [18] and 2D GL equation for superconductors [19].

In the present article we describe the state of a superconducting ring at temperature close to the critical one ($T_c$), when the thermal activation of phase slips (TAPS) is promoted. As discussed in Supplemental Material (SM), within a mean-field approach, this superconducting state can be modelled as composed by two kinds of particles: paired charge carriers, which do not provide resistance and possess phase coherence, and *quasiparticles* (QPs), which provide resistance and are phase incoherent. Hereafter, we refer to the paired charge carriers as *Cooper pairs* (CPs) even if dealing with high-critical-temperature superconductors, whose pairing mechanism is different from the BCS one, what does not affect the phenomenology [20] as compared to the conventional superconductors. The dynamics of the superconducting state is entirely governed by screening supercurrents because the magnetic field is supposed to be lower than the first critical field ($B_{c1}$).

Our goal is to show that the dependence of the measured resistance on the external magnetic flux is a consequence of the spatial sensitivity of voltage leads, which – being comparable with the connectors to the ring (the stubs shown in Fig.1) – operate like Hall sensors. Specifically, we prove that the onset of superconducting screening currents in the ring affects the transmission of CPs determining, by incompressibility of charge carriers, a change in the sensed transversal distribution of QPs, what results in a variation of differential voltage measured across the ring.

We show that the sinusoidal contribution in MRO is related to the interference of CPs, which does not imply any change in the amplitude of the currents of both types of particles CPs and QPs.

We have applied our results to an inspiring experiment [21], where the fluxoid quantization (FQ) is observed in a ultranarrow YBCO nanoring. The ring arms along with the stubs are so thin that the parabolic background is substantially absent (see below) enabling to focus on the sinusoidal contribution of MRO that presents oscillations at the frequency $\Phi_0 = h/2e$, which are one order of magnitude larger than those due to the Little–Parks effect [22]. Though supposed to occur due to vortex dynamics triggering the nanowires to the resistive state, they have remained quantitatively unexplained until now. Vortex dynamics is prevented for magnetic fields lower than $B_{C1}$. As a consequence of the narrowness of the wires realizing the mesoscopic ring [23], $B_{C1}$ is pretty larger than the magnetic field $B_p$ that induces a flux quantum (period) preventing the involvement of vortex dynamics to model the MRO in the range of measurements [24].

The article is organized as follows:
In section 2, we retrace the main concepts on the fluxoid quantization in one-dimensional rings. In section 3, the dependence of screening currents on the density of CPs is explained. In section 4, the order parameter of a finite-size mesoscopic ring is described within the Ginzburg-Landau (GL) theory. In section 5, we describe the interference-based mechanism capable to affect the superconducting currents flowing through the ring. In section 6, the connection between the interference and the MRO is exposed. In section 7, the main observable consequences of the model are presented and a discussion on normal rings is provided within the framework developed for superconducting ones. In section 8, the model is validated by tracing out the MRO acquired on a YBCO finite-size mesoscopic ring. In section 9, conclusions are drawn. In Supplemental Material, we give an explanation on the two kinds of particles induced by TAPS.

2. The fluxoid quantization in one-dimensional rings

We have developed a simplified model within the GL theory to account for the onset of screening currents in a finite-size mesoscopic ring as opposite to the 1D case [17], where the radius was a fixed parameter.

In a 1D superconducting ring of radius $r$, pierced by a magnetic flux of density $B$, there develops a persistent supercurrent $J_s = 2e\, n_s v_s$ expressed in terms of the elementary electron charge $e$, the CP density $n_s$ and the supercurrent velocity $v_s$. The quantum state describing the condensate of CPs is defined by the order parameter

$$\widetilde{\Psi}_{1D}(\vartheta) = \sqrt{n_s} \exp\left\{i\, 2\pi \frac{\Phi'(\vartheta)}{\Phi_0}\right\}, \quad (3)$$

where $\Phi_0 = h/2e$ is the magnetic flux quantum. The fluxoid [23]

$$\Phi'(\vartheta) = \int_{\gamma(\vartheta)} (\Lambda \bar{J}_s + \bar{A}) \cdot d\bar{r}(\vartheta') \quad (4)$$

depends on the azimuthal angle $\vartheta'$ of the arc of the circular path $\gamma(\vartheta') = r\,\vartheta'$ of screening currents described through the London vector potential $\Lambda \bar{J}_s$, where $\Lambda = 4\pi\lambda^2/c^2$, $c$ is the speed of light in vacuum and $\lambda$ is the magnetic field penetration depth. The external flux is expressed through the vector potential $\bar{A} = \Phi \frac{\bar{v}}{2\pi r}$, where $\Phi = B\pi r^2$ and $\bar{v}$ is the unitary vector tangent to the arc length $\gamma$ of radius $r$, oriented righthanded with respect to $\bar{B}$. The order parameter fulfils the property to conserve the number of CPs because $|\widetilde{\Psi}_{1D}|^2 = n_s$ and the supercurrent density $\bar{J}_s$ is obtained through the equation

$$\bar{J}_s = \frac{e\,\hbar}{i\,m^*}\{\widetilde{\Psi}^*\overline{\nabla}\widetilde{\Psi} - \widetilde{\Psi}\overline{\nabla}\widetilde{\Psi}^*\} - \frac{4\,e^2|\widetilde{\Psi}|^2}{m^*c}\bar{A} \quad (5)$$

where $m^*$ is the effective mass of the charge carrier [15]. Single-valuedness of the order parameter requires a $2\pi$-periodicity of the order parameter of eq. (3) as a function of the azimuthal angle implying that eq. (4) fulfils

$$\Phi'(2\pi) = \oint_{\gamma(2\pi)} (\Lambda \bar{J}_s + \bar{A}) \cdot d\bar{r}' = n\,\Phi_0 \quad (6)$$

where $, n = 0, \pm 1, \pm 2\ldots$. Hereafter we assume a circular symmetry of currents, which enables us to represent eq. (6) as

$$\Lambda J_s 2\pi r = n\Phi_0 - \Phi. \quad (7)$$

Because of the fixed radius, the only parameter to be quantized is the condensate velocity

$$v_s(\Phi) = \frac{\hbar}{4\,m^*r}\left(n - \frac{\Phi}{\Phi_0}\right), \quad (8)$$

in accordance with eq. (2).

### 3. The supercurrent density and its critical value

Electrical measurements are conducted through four points: a probe current $I_{in}$ is set in the source-drain channel and a differential voltage is measured. In Fig. 1, a sketch of the sample shows $I_{in,out}$ stubs, which deliver the input current into the ring and collect the output current from the ring, and voltage ones $V_{in,out}$.

When $T < T_c$ and $B < B_{c1}$, a net $I_c$ is established in the system. The amplitude of $I_c$ is measured monitoring the value of the input current $I_{in}$ till the nonzero measured differential voltage occurs $\Delta V = V_{out} - V_{in} \neq 0$. The current source feeds the sample with a current $I_{in} = \int_{\Sigma_I} \bar{J}_{in} \cdot d\bar{s}$ where $\Sigma_I$ is the cross-section of the stubs. Transport measurements via four points imply current conservation $I_{in} = I_{out}$ that if $I_{in} < I_c$, yields $I_{in} = I_s = \int_{\Sigma_I} \bar{J}_s \cdot d\bar{s}$. The question is: when $I_{in}$ is increased, which parameter of $J_s = 2e\,n_s\,v_s$ is modified? The velocity $v_s$ cannot be changed by any external parameter (otherwise an acceleration would be provided to CPs promoting the transition to the normal state). Hence, we infer that the increase of $I_{in}$ stimulates the growth of $n_s$ till the critical density $n_c(T,B)$ is reached, above which (namely, when $J_c = 2e\,n_c\,v_s$) the superconducting state is switched off. If the critical current density $J_c(T,B)$ depends on the maximum density of CPs $n_c(T,B)$ with $v_s(T,B) = $ const, we can employ the Silsbee criterion [15] relating the value of the critical current density to the critical magnetic field [25]

$$J_c = B_c/\mu_0\lambda \quad (9)$$

to get a new insight on $\lambda$ and the London parameter $\Lambda$. Indeed, according to eq. (9), the magnetic penetration depth $\lambda$ is related to the ratio of critical values of $J_c$ and $B_c$ and consequently depends on $n_c(T,B)$ and not on $n_s$. Therefore, it makes sense to express the London parameter as a function of $n_c$ in place of $n_s$ in $\lambda$, yielding $\Lambda = \mu_0\,\lambda(T)^2 = \frac{m^*}{n_c(T)e^2}$.

From the experimental point of view, the study of $I_c$ is operated via four points measurements, where the source-drain current is injected through normal leads carrying current at velocity $v_F$. Hence, although $v_F$ represents an upper limit for the superconducting state, CPs can sustain a velocity comparable with it.

### 4. Mean-field order parameter of a two-dimensional ring

In order to provide a new degree of freedom to eq. (7), we expand the discussion to a finite-size mesoscopic ring that is geometrically defined by a thickness (along the axis, which coincides with the direction of $B$) $d$, an inner radius $r_i$ and an outer one $r_o$. The sample is mesoscopic in the following sense. The magnetic field can penetrate the ring because the magnetic field penetration depth provided by the Pearl penetration depth [26] $\lambda_P = \lambda^2/d$ ($\lambda \gg d$), fulfils $\lambda > \Delta r = r_o - r_i$. Under this condition, the order parameter of a structure is found by sampling $\widetilde{\Psi}$ of the ring in local screening currents (LSCs), which circulate in one and the same direction tending to expel the external field. For the sake of clarity, in the following we assume the ring placed in the XY plane of a Cartesian coordinate system and the magnetic field is oriented along the Z axis. According to this choice the distribution of LSCs is reported in Fig.2(a), and the generic order parameter is expressed as

$$\widetilde{\Psi}(\bar{r}) = \sqrt{n_s} \prod_j \exp i \frac{2\pi}{\Phi_0} \int_{\gamma_{(\bar{r})j}} \left[ \Lambda \bar{J}_{s_j}(\bar{r}') + \bar{A}_j(\bar{r}') \right] \cdot d\bar{r}' \quad (10)$$

where the product runs over the index $j$ that labels the LSCs ($\bar{J}_{s_j} = 2e\, n_s \bar{v}_{s_j}(\bar{r}')$) pierced by a flux described through the vector potential $\bar{A}_j$. In agreement with the outcomes of the 1D ring case, the velocity $v_{s_j}$ is a function of the position $\bar{r}'$ to guarantee that adjacent LSCs do not cancel each other. Although eq. (10) is useful for computational procedures, we have introduced it to show how the boundaries related to $r_i$ and $r_o$ tend to define two currents, which, as depicted in Fig.2(a) flow in the opposite directions: one close to the inner radius has a paramagnetic orientation as opposed to another one near the outer radius that has a diamagnetic behavior. This insight suggests considering the screening current distribution as composed of two opposite supercurrents differing by the sign of $\bar{v}_s$. Although richer configurations of dia- and para-supercurrents could be taken into account [27], we assume the resulting current pattern reported in Fig.2(b), where a couple of currents flowing in the opposite directions produce a zero supercurrent at the circumference of the effective radius $\rho$ [28], $J_s(\rho) = 0$.

Therefore, taking into account that the developed supercurrents are a function of a CPs density $n_s$ that is homogeneously distributed over the whole 2D finite-size mesoscopic ring, we use the following reasonable Ansatz for the mean-field order parameter:

$$\widetilde{\Psi}_{2D} = \sqrt{n_s} e^{i\alpha_d} e^{-i\alpha_p} e^{i\varphi} \equiv \Psi_0 \Psi_\alpha \Psi_\varphi \quad (11)$$

where $\Psi_0 = \sqrt{n_s}$, $\Psi_\alpha = e^{i\alpha_d} e^{-i\alpha_p}$ accounts for the geometric phase giving rise to persistent currents [29] and $\Psi_\varphi = e^{i\varphi}$ is the phase due to the magnetic flux. In particular, the geometric phases $\alpha_{d,p} = \frac{2\pi}{\Phi_0} \oint_{\gamma_{d,p}} \Lambda \bar{J}_{d,p} \cdot d\bar{r}$ are labelled with subscripts $d, p$ to indicate dia- and para-supercurrents, correspondingly, and relative paths as defined below, whereas $\varphi = 2\pi \frac{\Phi}{\Phi_0} = \frac{2\pi}{\Phi_0} \oint_{\gamma_{eff}} \bar{A} \cdot d\bar{r}$ is the circulation of the vector potential over the circumference $\gamma_{eff}$ defined by the average radius $r_{avg} = (r_o + r_i)/2$. The phases $\alpha_{d,p}$ can be written assuming no variation of the modulus of screening current density along the radius:

$$\begin{cases} \alpha_p = \frac{2\pi}{\Phi_0} \int_{r_i}^{\rho} \frac{1}{\Delta r} \left( \oint_{\gamma_{(r')}} \Lambda \bar{J}_s \cdot d\bar{r} \right) dr' \\ \alpha_d = \frac{2\pi}{\Phi_0} \int_{\rho}^{r_o} \frac{1}{\Delta r} \left( \oint_{\gamma_{(r')}} \Lambda \bar{J}_s \cdot d\bar{r} \right) dr'. \end{cases} \quad (12)$$

The total geometric phase becomes

$$\alpha_d - \alpha_p = \frac{2\pi}{\Phi_0} \frac{\pi \Lambda J_s}{\Delta r} (r_o^2 + r_i^2 - 2\rho^2), \quad (13)$$

That enables us to express the condition of FQ in a finite-width ring in the form

$$\frac{\pi \Lambda J_s}{\Delta r} (2\rho^2 - r_o^2 - r_i^2) = n\Phi_0 - \Phi. \quad (14)$$

From this equation the effective radius is found as a function of $\Phi$

$$\rho(\Phi) = \sqrt{r_m^2 + \Delta r \lambda_q \left( n - \frac{\Phi}{\Phi_0} \right)}, \quad (15)$$

where $r_m^2 = (r_o^2 + r_i^2)/2$ is the mean squared radius of the ring and $\lambda_q = \frac{\hbar}{2\, p_{CP}} \eta$ contains the CP momentum $2\, m^* v_s = p_{CP}$ and $\eta = \frac{n_c}{n_s} = \frac{J_c}{J_s}$. The winding number $n$ in eq. (15) starts from zero and grows by a unity each time $\Phi = n\, \Phi_0$ so that the inequality $0 < \left| \frac{\Phi}{\Phi_0} - n \right| < 1$ is obeyed for each $n$. In its turn, $\rho(\Phi)$ reveals a quasi-linear $\Phi_0$-periodic sawtooth behavior oscillating in the range $\{r_m, \sqrt{r_m^2 - \Delta r \lambda_q}\}$. Thus, $\rho(\Phi)$ decreases in accord with the enhancement of the dia- supercurrent contribution over each flux growth interval $n\, \Phi_0 < \Phi < (n+1)\, \Phi_0$.

The effective radius $\rho(\Phi)$ is depicted in the upper panel of Fig. 2(c). The stronger the inequality $r_m^2 > \Delta r \lambda_q$ is, the more the $\rho(\Phi)$ behavior is close to linear within each flux growth interval $\Phi_0 < \Phi < (n+1)\, \Phi_0$. The oscillation of $\rho(\Phi)$ induces the oscillation of the boundary between para- from dia-supercurrents as depicted in the lower panel of Fig. 2(c). The effective radius exists when the argument of the root in eq. (15) is positive what is guaranteed by the inequality:

$$r_m^2 > \Delta r \lambda_q \rightarrow v_s > \frac{\Delta r}{4 r_m^2} \frac{\hbar}{\eta\, m^*}. \quad (16)$$

From eq. (16) we deduce that $v_s$ strictly relies on the ring size: the bigger $r_m$ is, the lower $v_s$ can be. We calculate the minimum velocity to fulfil the condition of FQ in a finite-size mesoscopic ring: with $m^* \sim m_e$ [30], $r_o = 200$ nm and $r_i = 150$ nm, eq. (16) yields $v_s > 200$ m/s.
Aiming to figure out the effects of eq. (15) on dia- and para-supercurrents, we substitute the explicit expression of $\rho(\Phi)$ in the integration limits of eqs. (12) obtaining

$$\begin{cases} \alpha_p = \frac{4\pi^2 \Lambda J_p\, r_{avg}}{\Phi_0} \\ \alpha_d = \frac{4\pi^2 \Lambda J_d\, r_{avg}}{\Phi_0}, \end{cases} \quad (17)$$

where $J_{p,d} = J_s \left[1 \pm \frac{\lambda_q}{r_{avg}}\left(\frac{\Phi}{\Phi_0} - n\right)\right]/2$.

It follows from eqs. (17), that the finite-size mesoscopic ring can be mapped onto a one-dimensional one of radius $r_{avg} = (r_o + r_i)/2$, in which two currents effectively flow in the opposite directions and fulfil the condition of FQ in the form

$$2\pi\Lambda[J_d(\Phi) - J_p(\Phi)]r_{avg} = \Phi - n\Phi_0. \quad (18)$$

Furthermore, starting from eq. (18), we argue that the effective magnetic flux can be calculated as $\Phi = B\pi r_{avg}^2$. The net supercurrent density swirling in the ring $J_{d,ring} = J_d - J_p$ as a function of the flux is obtained from eq. (18) or, alternatively, by substituting eq. (11) in eq. (5):

$$J_{d,ring} = J_d - J_p = J_s \frac{\lambda_q}{r_{avg}}\left(\frac{\Phi}{\Phi_0} - n\right) = \frac{\Phi_0}{2\pi\Lambda\, r_{avg}}\left(\frac{\Phi}{\Phi_0} - n\right). \quad (19)$$

Here we use the gradient operator in the form $\nabla = \partial_\theta / r_{avg}$ that provides $\nabla(\alpha_d - \alpha_p) \propto J_d - J_p$. Eq. (19), developed in a 2D ring resembles the same result achieved in the 1D system, (see eq. (7)), provided the average radius in place of the fixed one.

5. The transmission function

The transmission of supercurrent across the ring is modulated by an interference mechanism that can be expressed through the equation (7) of [24] relating the order parameter variation affecting the transmission through the ring

$$\widetilde{\Psi}_{out} = T(\Phi)\widetilde{\Psi}_{in}, \quad (20)$$

where $T(\Phi)$ is an even function of the magnetic flux [11] obtained by calculating the phase accumulated by supercurrent splitting into the two arms of the ring:

$$\widetilde{\Psi}_{out} = \frac{1}{2}\widetilde{\Psi}_{in}\left\{\exp\left[i\frac{2\pi}{\Phi_0}\pi\Lambda(J_d - J_p)r_{avg}\right] + \exp\left[-i\frac{2\pi}{\Phi_0}\pi\Lambda(J_d - J_p)r_{avg}\right]\right\} =$$
$$= \widetilde{\Psi}_{in}\cos\left\{\frac{\pi}{\Phi_0}2\pi\Lambda(J_d - J_p)r_{avg}\right\} =$$
$$(-1)^n\,\widetilde{\Psi}_{in}\cos\left\{\pi\frac{\Phi}{\Phi_0}\right\}. \quad (21)$$

In the last equality, eq. (18) is used. Hence, as a consequence of eq. (6), the transfer function for a finite-size mesoscopic ring is

$$\frac{\widetilde{\Psi}_{out}}{\widetilde{\Psi}_{in}} = T(\Phi) = (-1)^n \cos\pi\frac{\Phi}{\Phi_0}, n = 0, \pm 1, \pm 2\ .... \quad (22)$$

The relation between input and output supercurrent density is achieved by substituting eq. (22) in eq. (5) yielding

$$J_{s,out} = J_{s,in}\cos^2\pi\frac{\Phi}{\Phi_0}. \quad (23)$$

The above dependence of supercurrent density on $\Phi$ agrees with the oscillations detected through current–voltage characteristics $IV(\Phi)$ acquired in both finite-size mesoscopic rings [24] and SQUIDs [31]. However, it is fundamental to realize that eq. (23) involving the transmission of supercurrent density, takes sense only *locally*, because the *total* injected supercurrent is conserved in the experiment as discussed in the next section.

6. The resistance as a function of the external magnetic field

In order to figure out the implications of eq. (23) in MRO, we have to realize how the modulation of the transmitted supercurrent density meets the experimental fact that the total current is conserved. Indeed, four-points measurements are conducted by setting the input current $I_{in}$, so that the equality

$$I_{in} = I_{out} \quad (24)$$

is obeyed in experiment.

Further insights to grab the meaning of eqs. (23) and (24) are achieved if we briefly introduce the standard experimental procedure to acquire MRO. First, the sample state is prepared by setting the temperature slightly below the critical one [22] $T \sim T_c$, where $T_c$ represents the thermodynamical critical temperature [32] found from the superconducting-to-normal transition. Under this experimental condition, the $IV$ characteristic shows a zero-bias resistance as sketched in Fig. 2(d). The origin of the zero-bias resistance state is the thermal activation of phase slips (TAPS) [33] [34] [35]. Owing to the residual density of CPs, phase coherence of the state is preserved as long as $I_{in}$ is weaker than the residual critical one identified by the kinks $I_c^T$ in the *IV* characteristic. Indeed, the inequality $I_{in} < I_c^T$ prevents activation of vortex motion tending to blur away the MRO [24] [36]. MRO are recorded by monitoring the change of the voltage difference $\Delta V = V_{out} - V_{in}$ through a lock-in amplifier referenced with the phase of a low-frequency (~10 Hz) ac zero-bias excitation current labelled as $I_{in}(MR)$ in Fig. 2(d). After a fast transient, each value $\Delta V(B)$ is constant during experiment, revealing that the observed dynamics of the persistent current is stationary.

The zero-bias resistance is modelled through the presence of a mixture of QPs (*qp*) and CPs (*s*), so that eq. (24), measured by an ammeter (*A*), turns into

$$I_{in,qp}^{(A)} + I_{in,s}^{(A)} = I_{out,qp}^{(A)} + I_{out,s}^{(A)}. \quad (25)$$

Since there are no traps for charge carriers or CPs in the system, the following equations hold true:

$$\begin{cases} I_{in,qp}^{(A)} = I_{out,qp}^{(A)}, \\ I_{in,s}^{(A)} = I_{out,s}^{(A)}. \end{cases} \quad (26)$$

In order to represent the eq. (25) in terms of the differential voltage $\Delta V$, a discussion on the Ohm's law in mesoscopic superconducting devices is outlined in what follows.

The dependence of $\Delta V$ on the input current $I_{in}$ cannot be assumed as $\Delta V = R\, I_{in}$, because the differential voltage is sensitive to QPs dynamics only. Yet the observed sinusoidal $\Delta V$ as a function of $B$ cannot be ascribed to any change of either CPs or QPs currents because according to eq. (26) they are conserved separately.

If a small magnetic field creating a few flux quanta and a multiple of several values $B_p = \Phi_0/\pi r_{avg}^2$ would be capable to increase the number of QPs, damped oscillation of the resistance would occur, because $B_p \sim B_c$, but this phenomenon is observed in neither conventional, [37] [38] nor high-critical-temperature superconductors [22] [39]. Another interpretation of MRO is based on the vortex flow modulation [39] driven by the oscillations of CPs velocity. This interpretation has to be rejected as well because the first critical field, which scales as $B_{c1} \sim \Phi_0/w^2$ [23], where $w$ is the width of the mesoscopic wires (stubs included) composing the sample, usually satisfies the inequality $B_{c1} > B_p$. Hence, the first oscillations of MR cannot be governed by any vortex dynamics, because vortices are nucleated in the structure only when the magnetic field exceeds several values of $B_p$. As shown in [24], vortices can play a role in the formation of the MRO background provided that they remain trapped within the Gibbs barrier of mesoscopic wires [40].

Instead, the sinusoidal trend of MRO suggests that the phenomenon completely depends on the order-parameter phase rather than on its modulus. Hence, as long as no new QPs are generated by the magnetic field $B < B_{c1}$, the mechanism guiding MRO refers to a change of the QPs sensed by the voltage leads, so that the relation between the differential voltage and the QP current is as follows: $\Delta V(T,B) \sim R(T)\Delta I_{qp}(B)$, where $R(T)$ is the value of the resistance at the temperature of the MRO and $\Delta I_{qp}(B)$ is related to the change of the edge density of QPs between the input and output stubs (see Fig. 1). While voltage leads fabricated on top of stubs [41] may lack sensitivity to QPs moving inside the stubs, voltage leads realized at one stub's edge may lack sensitivity to QPs moving on the opposite stub edge (see Fig.1 in [36]). The sensitivity areas are depicted as shadowed regions close to voltage leads in Figs.1 and 3.

Hence the current conservation expressed in eq. (24) can be translated in terms of voltage by integrating the current density of all charge carriers $J_{in/out,qp/s}$ over a surface ($\Sigma_V$) in proximity of the voltage leads area. This sensitivity surface is parallel to $\bar{y}$ (see Fig. 2(b)) and smaller than the whole stub cross-section ($\Sigma_V < \Sigma_I$). In this way, we get instead of eq. (25)

$$I_{in,qp}^{(V)} + I_{in,s}^{(V)} = I_{out,qp}^{(V)} + I_{out,s}^{(V)}, \quad (27)$$

while the equalities in eqs. (26) do not hold when current densities are integrated over the surfaces relative to the voltage sensitivity areas ($\Sigma_{V_{in,out}}$). The eq. (27) is well justified because all charge carriers move in the same matrix of "sites" realized by the crystal lattice independently no matter if the conditions favor the pairing. Hence, if the charge carrier flow is uncompressible, the number of charges within two sections of sensitivity areas is conserved $\int_{\Sigma_{V_{in}}} (\bar{J}_{qp} + \bar{J}_s) \cdot d\bar{\Sigma} = \int_{\Sigma_{V_{out}}} (\bar{J}_{qp} + \bar{J}_s) \cdot d\bar{\Sigma}$. As discussed in section 3, the verification of eq. (27) is also sustained by the experimental evidence that the *transport* velocity of both type of particles fulfils $\frac{dz}{dt} = v_F = \text{const}$.

The current conservation enables to measure the effects of FQ on CPs transmission through the change of QPs distribution according to voltage leads:

$$I_{in,s}^{(V)} - I_{out,s}^{(V)} = I_{out,qp}^{(V)} - I_{in,qp}^{(V)}. \quad (28)$$

In order to quantify the two terms of eq. (28), we have to recall that the distribution of CPs and QPs flowing through the ring is defined by the dynamics of the effective radius $\rho(\Phi)$ [eq. (15)]. With this purpose, we sketch the dependence of CP and QP paths as a function of $B$ in Fig.3 (yellow and green arrows represent currents of QPs and CPs, respectively). For $B = 0$ [Fig.3(a)], $I_{in}$ flows through the ring embedding a homogeneous mixture of CPs and QPs. When $B \neq 0$ [Fig. 3(b)] the motion of CPs is conditioned by the onset of dia- and para-supercurrents, whose distribution is related to the dynamics of $\rho(\Phi)$. As seen from Fig.3(b), CPs can propagate as either the dia- or para-supercurrents, which tend to adjoin the edges of the ring. Instead, within this simplified scenario, QPs propagate in between dia- and para-supercurrents, where $J_s$ is nominally zero. Therefore, the dynamics of $\rho(\Phi)$ induces the path of QPs to oscillate as a function of $B$, thus producing a change in the sensed local QP density in the vicinity of the $V_{in}$ and $V_{out}$ leads.

The integration of eq. (23) over $\Sigma_V$ within the voltage sensitivity volumes to get the left side of (28) leads to

$$I_{in,s}^{(V)} - I_{out,s}^{(V)} = I_{in,s}^{(V)} \sin^2 \pi \frac{\Phi}{\Phi_0}. \qquad (29)$$

In its turn, the right side of eq. (28) can be expressed using the Ohm's law:

$$I_{out,qp}^{(V)}(B) - I_{in,qp}^{(V)} = \frac{\Delta V_{int}(B) + \Delta V_0(T)}{R(T)}, \qquad (30)$$

where $\Delta V_{int}(B)$ is the interference-dependent, sinusoidal in the magnetic flux contribution, $\Delta V_0(T)$ is the background differential voltage at $\Phi = 0$ and $R(T) \equiv \Delta V_0(T)/I_{in}$ is the resistance of the sample acquired from the normal-to-superconducting transition without a magnetic field [24].

Substituting eqs. (29) and (30) into eq. (28), we arrive at

$$\Delta V_{int}(B) = R(T) I_{in,s}^{(V)} \sin^2 \pi \frac{\Phi}{\Phi_0} + \Delta V_0(T) \qquad (31)$$

To achieve the experimentally detectable MR $R(T,B) \equiv \frac{\Delta V_{int}}{I_{in}}$, both sides of eq. (31) are divided by $I_{in}$, what results in

$$R(T,B) = R(T) \left[ 1 + \frac{I_{in,s}^{(V)}}{I_{in}} \sin^2 \pi \frac{\Phi}{\Phi_0} \right]. \qquad (32)$$

It is worth noting that the parameter $I_{in,s}^{(V)} \leq I_c^T$ in eq. (32) is the only parameter to fit MROs, because $R(T)$ is acquired from the normal-to-superconducting transition and $\Phi = B\pi r_{avg}^2$ is determined by geometry of the sample.

Since $I_{in}$ is a homogeneous superposition of CP and QP currents, the ratio between $I_{in,s}^{(V)}$ and $I^{(A)}{}_{in,s}$ scales like the cross-section areas yielding $\frac{I_{in,s}^{(V)}}{\Sigma_V} = \frac{I_{in,s}^{(A)}}{\Sigma_I}$, wherefrom $I_{in,s}^{(V)} = \frac{\Sigma_V}{\Sigma_I} I_{in,s}^{(A)}$. Therefore, in the fitting procedure the inequalities $I_{in,s}^{(A)} \leq I_{in}$ and $\frac{\Sigma_V}{\Sigma_I} < 1$ should be fulfilled.

It is important to observe that in eq. (32) for a fixed $T$, the supercurrent in the prefactor of the sinusoidal trend in the MRO, either read by the ammeter $I_{in,s}^{(A)}$ or by the voltmeter $I_{in,s}^{(V)} = \frac{\Sigma_V}{\Sigma_I} I_{in,s}^{(A)}$, does not depend on $B$. This property differs from the widely used concept in quantum rings based on the oscillations of $I_c(B)$, which are extracted from $IV(B)$ characteristics by applying an *ad hoc* voltage threshold criterion. The present quantum-interference-based model attributes the cosinusoidal modulation of supercurrents only to $J_{s,out}(\Phi)$ [eq. (23)], and it gets the resulting sinusoidal trend in the MRO [eq. (32)] while keeping $|\widetilde{\Psi}_{2D}| =$ const.

7. Discussion

According to eq. (32), the dependence of MRO on $T$ and $B$ is factorized, so that the temperature dependence is included through the resistance $R(T)$ only, the flux dependence is determined by the geometry of the sample and the ratio between the total input current and its superconducting constituent.

In our model, the periodic trend in MR as a function of the magnetic field is achieved when keeping $|\Psi_{2D}(T,B)|^2 =$ const. The periodic modulation of MR does not translate an oscillation of QP density, as distinct from the LP effect [17]. Instead, the parabolic background of MR acquired on conventional superconductors, as in the original Little and Parks experiment [13], is certainly a consequence of the growth of QP current because $B_p \sim B_c$.

In conventional superconductors, the sinusoidal contribution to MR can be damped, and it can be modelled with the presented approach provided that the CP density is accounted through a function $n_s(T,B) = n_{s0}(T) e^{-\left(\frac{B}{B_c}\right)^2}$ as shown in eq. (10) of ref. [24].

In non-superconducting rings, the mechanism fostering MRO is related to the single-valuedness of the wave function of phase coherent charge carriers that according to eq. (1) leads to

$$\frac{1}{h} \oint_{\gamma[\bar{p}(\bar{r})]} \bar{p} \cdot d\bar{r} = \frac{\Phi}{\Phi_0} - n\,; \quad n = 0, \pm 1, \pm 2 \ldots. \qquad (33)$$

This equation, where $\Phi_0 = h/e$, is a version of the FQ for unpaired charge carriers. The external flux $\Phi$ sets a relation between the charge carrier momentum and the closed path in the ring bounded by radii $r_i$ and $r_o$, so that within a quasiclassical approach, a charge carrier of momentum $\bar{p}$ circulates over a closed trajectory $\gamma[\bar{p}(\bar{r})]$. In the same way as the superconducting case, quantum interference in normal mesoscopic rings can be modelled by assuming the flow of two types of charge carriers: those preserving or not phase memory. Only those preserving phase memory during the transport will satisfy eq. (33) following paths defined by specific radii accordingly to charge carrier velocities distributed around the Fermi velocity.

In the same fashion as in the superconducting case, which is considered in detail in the present paper, the interference governs the closed trajectories of charge carriers, which create a change in the distribution of charge carriers between $V_{in}$ and $V_{out}$, resulting in an oscillating MR as a function of the magnetic field.

The model presented in this paper, offers a way to interpret the observed phenomena Ref. [42] related to quantum jumps in atomic-like systems realized with superconducting nanorings of Al. Having extended the theory by Tinkham [17] to model the critical currents dynamics in nanorings with both asymmetric arms and a link-up connection, the authors of Ref. [42] wonder why the oscillation of the critical currents versus $B$ in nanorings with asymmetric arms does not exhibit jumps contrary to the case of the nanorings with a link-up connection. According to our model, these anomalies can be straightforwardly understood. In the case of nanorings with asymmetric arms, just a deformation with respect to the usual sinusoidal behavior is expected because of an additional phase contribution due to some extra supercurrent in the wider arm. As distinct from that, jumps in nanorings with a link-up connection can arise from the discretization of the circular paths, which force the effective radius and the relative para- to dia-supercurrents distribution to undergo an abrupt change at certain specific values of the magnetic field.

8. Model validation

In order to validate the model of MRO reported in eq. (32), we employ data of Ref. [21], where the MRO of a YBCO finite-size mesoscopic ring are investigated in the vicinity of $T_c$. The sample under investigation featured by an inner/outer radii $r_i/r_o = 150nm/200nm$ and a stub width of about $w = 50$ nm. The narrowness of the sample guarantees that the first critical field fulfils $B_{c1} \sim \Phi_0/w^2 \gg \Phi_0/\pi r_{avg}^2$, so that nucleation of vortices is activated after several flux quanta have pierced the ring. The absence of vortices prevents the onset of the parabolic background in MRs [24] making these curves available for testing the model developed.

In Fig. 4, black curves represent experimental MRs for three temperatures whereas red curves are obtained from eq. (32) with $r_{avg} = 170$ nm, $R(T)$ ranges between 2.36 Ω and 0.46 Ω, whereas $\frac{I_{in,s}^{(V)}}{I_{in}}$ ranges from 0.29 to 0.19 as the temperature decreases. The present model perfectly describes the experimentally detected MRO of Ref. [21].

The fact that FQ induces a redistribution of CPs/QPs detectable by voltage leads is further proven in Ref. [43] where a Hall bar is used to measure the magnetic moment of mesoscopic structures made of aluminum. The sample magnetization shows a damped saw-tooth trend as a function of the magnetic field, that is a consequence of the oscillations of the screening current modulation [eq. (19)]. Its amplitude lowers after each oscillation because of the comparability between $B_p$ and $B_c$ determining the reduction of both the CP density and supercurrents.

9. Conclusions

We have presented a model based on the GL theory to account for the sinusoidal MRO in superconducting finite-size mesoscopic rings either made of low- and high-critical temperature-superconductors. As distinct from the previous approaches, the magnetic field does not affect the number of scattering particles (QPs). Acting on the phase of the order parameter, it induces the sinusoidal variation of sensed differential voltage as a consequence of the change in the local distribution of CPs. In other words, the proposed transmission model accounts for the sinusoidal contribution of a MR by retaining $\left|\widetilde{\Psi}_{2D}(B)\right|^2 = $ const. In particular, we have found the mean-field order parameter $\widetilde{\Psi}_{2D}$ by averaging the para- and dia-supercurrents driven by system boundaries. Fluxoid quantization is granted by a specific distribution of dia- and para-supercurrents that results in an oscillatory shift of the effective radius $\rho$, where the screening current is nominally zero. Hence, the sinusoidal in the magnetic flux MRO result from a modification of input/output CPs and QPs trajectories as a consequence of supercurrent dynamics in the finite-size ring. This effect modifies the number of QPs sensed by voltage leads by a quantity $\Delta V_{int}$, which is not a function of a change of the QPs and CPs currents, but is rather related to a change of the local current densities of QPs and CPs in the vicinity of the voltage leads.

The magnetic flux breaks the symmetry of currents between input and output stubs by stimulating an order in supercurrents circulating in the ring. This order consists in a spatial separation of supercurrents from the dissipative currents. The ideal separation of dia- and para-supercurrents is just a consequence of the model considering a fluid, in which the densities of CPs and QPs are set by the probability of a phase slip event over the surface of the sample. As properly discussed in supplemental material (S1), TAPS induce an ongoing mutual transformation between CPs in QPs, so that the real appearance of the currents, instead of Fig.3(b), would consist of smeared bands of dia- and para-supercurrents and the QP channel.

The presented model and related outcomes on the measurable interference-dependent voltage are of direct value for other devices driven by coherence, like Josephson junctions or nanopatterned superconductor films.

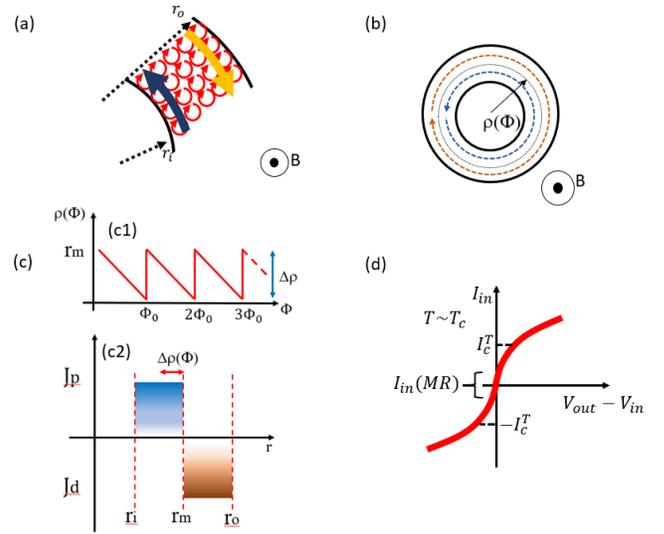

Fig. 2. (a) Slice of a finite/width ring showing the onset of diamagnetic LSCs depicted by the small circular arrows. Due to the presence of inner ($r_i$) and outer ($r_o$) radii LSCs envelope in a paramagnetic current (blue arrow) adjacent to $r_i$ and a diamagnetic (orange arrow) supercurrent adjacent to $r_o$. (b) Mean-field representation of total supercurrents showing paramagnetic (blue arrow) and diamagnetic behavior (red) separated by a circular path set by the effective radius $\rho$, where the resulting supercurrent is nominally zero. (c): (c1) Oscillation of the effective radius as a function of the flux quantum piercing the ring. (c2) Schematic representation of the para- and dia-supercurrent distribution over the ring radius. The double red arrow indicates the tuning of the effective radius $\rho$ as a function of the magnetic flux $\Phi$. See the text for explanation.
(d) Typical $IV$ characteristic of a superconducting sample close to $T_c$. $I_{in}$(MR) represents the zero-bias input current, whereas $I_c^T$ labels the residual critical current when the superconducting state is affected by TAPS.

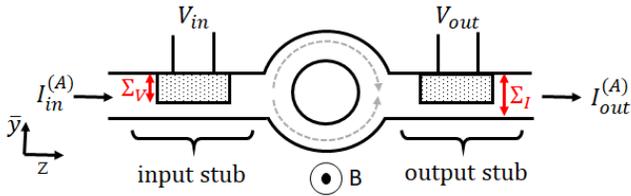

Fig. 1. Sketch of a finite-size mesoscopic ring realized for four-points measurements of the current–voltage characteristics.

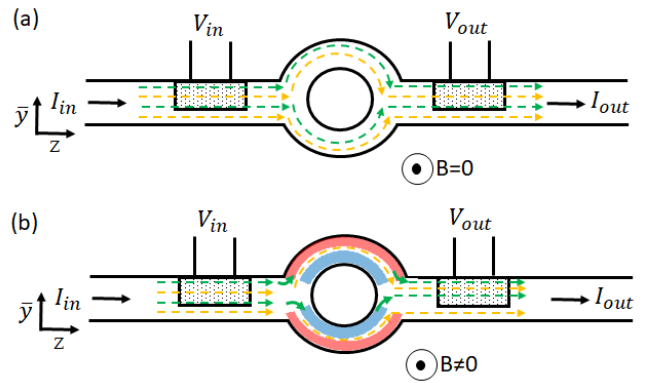

Fig. 3. (a) Schematic representation of input currents of QPs (yellow) and CPs (green) for $B = 0$. In the absence of an external magnetic flux, the incoming and outgoing currents are provided by a homogeneous mixture of CPs and QPs. (b) The presence of the magnetic field stimulates the onset of dia- (red band) and para- (azure band) supercurrents, which guide the flow of CPs through the ring. The flow of QPs is enabled in between dia- and para-supercurrents, where $J_s = 0$. See the text for explanation.

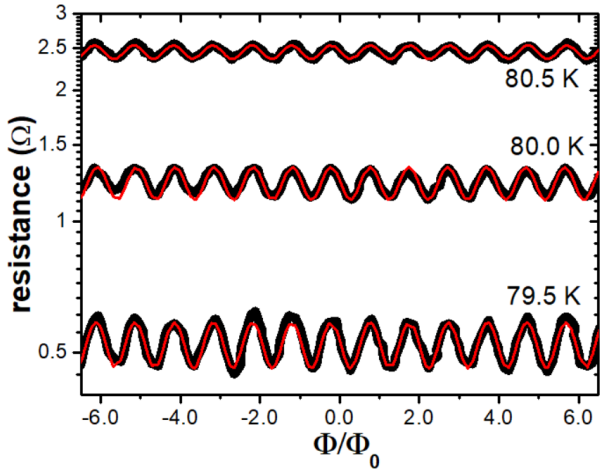

Fig. 4. Comparison between the experimental MRs of a YBCO finite-size mesoscopic ring (After [21]) and the theoretical result eq. (32) of the present model, represented by black and red curves, respectively. In [21] (see Fig. 1), two mesoscopic rings were investigated: the first one had arms and stubs as narrow as about 50nm, while the second ring had wide arms and stubs of the order of 150nm. MRO presented here relate to the narrower sample. The properties of wider mesoscopic rings were already discussed in [24].

Funding sources and acknowledgments. Content in the funding section will be generated entirely from details submitted to Prism.

Conflict of interests: Authors declare no conflicts of interests

Acknowledgments. The present work has been supported by the German Research Foundation (DFG) project #FO 956/6-1 and by the COST Action CA16218 (NANOCOHYBRI) of the European Cooperation in Science and Technology. V. M. F. acknowledges a partial support by the National Research Nuclear University "MEPhI" (Moscow, Russia) and ZIH TU Dresden (Germany).

References

[1] E. O. Göbel and U. Siegner, *The New International System of Units (SI): Quantum Metrology and Quantum Standards* (Wiley, 2019).

[2] S. Gazibegovic, D. Car, H. Zhang, S. C. Balk, J. A. Logan, M. W. A. De Moor, M. C. Cassidy, R. Schmits, D. Xu, G. Wang, P. Krogstrup, R. L. M. Op Het Veld, K. Zuo, Y. Vos, J. Shen, D. Bouman, B. Shojaei, D. Pennachio, J. S. Lee, P. J. Van Veldhoven, S. Koelling, M. A. Verheijen, L. P. Kouwenhoven, C. J. Palmstrøm, and E. P. A. M. Bakkers, *Epitaxy of Advanced Nanowire Quantum Devices*, Nature 548, 434 (2017).

[3] H. L. Huang, D. Wu, D. Fan, and X. Zhu, *Superconducting Quantum Computing: A Review*, Sci. China Inf. Sci. 63, 1 (2020).

[4] V. M. Fomin, *Physics of Quantum Rings* (Springer International Publishing, 2018).

[5] S. Datta, *Electronic Transport in Mesoscopic Systems* (Cambridge University Press, 1995).

[6] S. Washburn, *Aharonov-Bohm Effects in Loops of Gold*, in *Mesoscopic Phenomena in Solids*, edited by B. L. Altshuler, P. A. Lee, and R. A. Webb (Elsevier Science Publishers B.V., 1991, 1991).

[7] S. Washburn and R. A. Webb, *Aharonov-Bohm Effect in Normal Metal Quantum Coherence and Transport*, Adv. Phys. 35, 375 (1986).

[8] M. Büttiker, *Four-Terminal Phase-Coherent Conductance*, Phys. Rev. Lett. 57, 1761 (1986).

[9] V. Chandrasekhar, M. J. Rooks, S. Wind, and D. E. Prober, *Observation of Aharonov-Bohm Electron Interference Effects with Periods h/e and h/2e in Individual Micron-Size, Normal-Metal Rings*, Pis'ma Zh. Eksp. Teor. Fiz 35, 476 (1982).

[10] V. M. Fomin, *Self-Rolled Micro- and Nanoarchitectures* (De Gruyter, 2021).

[11] N. Byers and C. N. Yang, *Theretical Consideration on Concerning Quantized*

*Magnetic Flux in Superconducting Cylinders*, Phys. Rev. Lett. 7, 46 (1961).

[12] W. A. Little and R. D. Parks, *Observation of Quantum Periodicity in the Transition Temperature of a Superconducting Cylinder*, Phys. Rev. Lett. 9, 9 (1962).

[13] R. D. Parks and W. A. Little, *Fluxoid Quantization in a Multiply-Connected Superconductor*, Phys. Rev. 133, A97 (1964).

[14] M. Tinkham, *Consequences of Fluxoid Quantization in the Transitions of Superconducting Films*, Rev. Mod. Phys. 36, 268 (1964).

[15] M. Tinkham, *Introduction to Superconductivity* (McGrow-Hill, Inc., 1996).

[16] J. T. Devreese, V. M. Fomin, V. N. Gladilin, and J. Tempere, *Oscillatory Persistent Currents in Quantum Rings: Semiconductors versus Superconductors*, Phys. C Supercond. Its Appl. 470, 848 (2010).

[17] M. Tinkham, *Effect of Fluxoid Quantization on Transitions of Superconducting Films*, Phys. Rev. 129, 2413 (1963).

[18] L. Wendler, V. M. Fomin, and A. A. Krokhin, *Relation between Persistent Current and Band Structure of Finite-Width Mesoscopic Rings*, Phys. Rev. B 50, 4642 (1994).

[19] V. Fomin, V. Misko, J. Devreese, and V. Moshchalkov, *Superconducting Mesoscopic Square Loop*, Phys. Rev. B - Condens. Matter Mater. Phys. 58, 11703 (1998).

[20] E. H. Brandt, *The Flux-Line Lattice in Superconductors*, Reports Prog. Phys. 58, 1465 (1995).

[21] R. Arpaia, S. Charpentier, R. Toskovic, T. Bauch, and F. Lombardi, *YBCO Nanorings to Probe Fluxoid Quantization in High Critical Temperature Superconductors*, Phys. C Supercond. Its Appl. 506, 184 (2014).

[22] F. Carillo, G. Papari, D. Stornaiuolo, D. Born, D. Montemurro, P. Pingue, F. Beltram, and F. Tafuri, *Little-Parks Effect in Single Nanoscale $YBa_2Cu_3O_{6+x}$ Rings*, Phys. Rev. B 81, 054505 (2010).

[23] G. Stan, S. B. Field, and J. M. Martinis, *Critical Fields for Vortex Expulsion from Narrow Superconducting Strips*, Phys. Rev. Lett. 92, 097003 1 (2004).

[24] G. P. Papari and V. M. Fomin, *Interplay between the Quantum Interference and Current Localization Phenomena in Superconductor Non-Ideal Mesoscopic Rings*, Supercond. Sci. Technol. 32, (2019).

[25] C. P. J. Poole, editor , *Handbook of Superconductivity* (Academic Press, 2000).

[26] J. Pearl, *Current Distribution in Superconducting Films Carrying Quantized Vortices*, Appl. Phys. Lett. 5, 65 (1964).

[27] H. Zhao, V. M. Fomin, J. T. Devreese, and V. V Moshchalkov, *A New Vortex State with Non-Uniform Vorticity in Superconducting Mesoscopic Rings*, Solid State Commun. 125, 59 (2003).

[28] S. Yampolskii, F. Peeters, B. Baelus, and H. Fink, *Effective Radius of Superconducting Rings and Hollow Cylinders*, Phys. Rev. B 64, 1 (2001).

[29] V. M. Fomin, editor , *Physics Of Quantum Rings*, 2nd Editio (Springer International Publishing, 2018).

[30] J. E. Hirsch, *What Is the Speed of the Supercurrent in Superconductors?*, Arxiv : 1605.09469v4 15 (2016).

[31] A. Barone and G. Paterno', *Physics and Applications of Josephson Effect* (John Wiley & Sons, 1982).

[32] A. Bezryadin, *Quantum Suppression of Superconductivity in Nanowires*, J. Phys. Condens. MATTER 20, 1 (2008).

[33] V. Ambegaokar and B. I. Halperin, *Voltage Due to Thermal Noise in the Dc Josephson Effect*, Phys. Rev. Lett. 22, 1364 (1969).

[34] D. E. McCumber and B. I. Halperin, *Time Scale*

of Intrinsic Resistive Fluctuations in Thin Superconducting Wires*, Phys. Rev. B 1, 1054 (1970).

[35] J. S. Langer and V. Ambegaokar, *Intrinsic Resistive Transition in Narrow Superconducting Channels*, Phys. Rev. 164, 498 (1967).

[36] G. P. Papari, A. Glatz, F. Carillo, D. Stornaiuolo, D. Massarotti, V. Rouco, L. Longobardi, F. Beltram, V. M. Vinokur, and F. Tafuri, *Geometrical Vortex Lattice Pinning and Melting in YBaCuO Submicron Bridges*, Sci. Rep. 6, 1 (2016).

[37] M. Morelle, V. Bruyndoncx, R. Jonckheere, and V. Moshchalkov, *Critical Temperature Oscillations in Magnetically Coupled Superconducting Mesoscopic Loops*, Phys. Rev. B 64, 1 (2001).

[38] O. J. Sharon, A. Shaulov, J. Berger, A. Sharoni, and Y. Yeshurun, *Current-Induced SQUID Behavior of Superconducting Nb Nano-Rings*, Sci. Rep. 6, 2 (2016).

[39] I. Sochnikov, A. Shaulov, Y. Yeshurun, G. Logvenov, and I. Božović, *Oscillatory Magnetoresistance in Nanopatterned Superconducting La1.84Sr0.16CuO4 Films*, Phys. Rev. B 82, 094513 (2010).

[40] K. H. Kuit, J. R. Kirtley, W. van der Veur, C. G. Molenaar, F. J. G. Roesthuis, A. G. P. Troeman, J. R. Clem, H. Hilgenkamp, H. Rogalla, and J. Flokstra, *Vortex Trapping and Expulsion in Thin-Film YBCO Strips*, Phys. Rev. B 77, 134504 (2008).

[41] W. G. van der Wiel, Y. V. Nazarov, S. De Franceschi, T. Fujisawa, J. M. Elzerman, E. W. G. M. Huizeling, S. Tarucha, and L. P. Kouwenhoven, *Electromagnetic Aharonov-Bohm Effect in a Two-Dimensional Electron Gas Ring*, Phys. Rev. B - Condens. Matter Mater. Phys. 67, 1 (2003).

[42] V. L. Gurtovoi, A. I. Il'in, and A. V. Nikulov, *Experimental Investigations of the Problem of the Quantum Jump with the Help of Superconductor Nanostructures*, Phys. Lett. Sect. A Gen. At. Solid State Phys. 384, (2020).

[43] M. Morelle, J. Bekaert, and V. Moshchalkov, *Influence of Sample Geometry on Vortex Matter in Superconducting Microstructures*, Phys. Rev. B 70, 094503 (2004).

[44] G. Papari, F. Carillo, D. Stornaiuolo, L. Longobardi, F. Beltram, and F. Tafuri, *High Critical Current Density and Scaling of Phase-Slip Processes in YBaCuO Nanowires*, Supercond. Sci. Technol. 25, 35011 (2012).

[45] W. A. Little, *Decay of Persistent Currents in Small Superconductors*, Phys. Rev. 156, 396 (1967).

[46] M. Tinkham and C. N. Lau, *Quantum Limit to Phase Coherence in Thin Superconducting Wires*, Appl. Phys. Lett. 80, 2946 (2002).

Supplemental Material:

Thermal activation of phase slips as a source of partition between Cooper pairs and quasiparticles

The normal-to-superconducting transition of a superconducting sample is usually traced by monitoring the resistive transition $R(T)$ that −below a certain temperature $T_c$ −shows an exponential decay down to a nominally zero-resistance state [15]. The reference model for describing the normal-to-superconducting transition is based on the GL theory and invokes the onset of a barrier of the free energy $\Delta F$ that preserves the macroscopic phase coherence by preventing the order parameter to escape from it. The higher $\Delta F$ is, the lower is the probability of an event of phase slip promoted by a thermal excitation (TAPS), whose amplitude scales as $k_B T$, where $k_B$ is the Boltzmann constant. A phase slip means that the superconducting state loses phase coherence becoming normal for a time of the order of the GL time $\tau_{GL} \leq 10\text{ns}$ [34] over a surface of the order of the squared coherence length $\xi^2$ that is worth about $4\text{nm}^2$ for YBCO [44].

The experimental manifestation of TAPS is the broadening of the $R(T)$ [44] caused by the presence of local phase slips disabling the onset of a net superconducting flow for $T < T_c$ as opposite to what happens in bulk materials characterized by a sharp $R(T)$ because the presence of TAPS does not prevent a net supercurrent at $T \leq T_c$.

According to the theory by Little [45] [32] the resistive transition in nanowires can be modelled as

$$R(T) = R_0 \, e^{-\frac{\Delta F(T)}{k_B T}} \quad (S1)$$

where $R_0$ is the value of the resistance at $T_c$. The exponential function tunes the probability [32] of a phase slip event from $T_c$, where $\Delta F(T_c) = 0$, by means of the ratio between the free energy barrier and the thermal energy. The eq. (S1) accounts for a phase slip event that can happen everywhere over the nanowire. Its rate is weighted by the onset of a superconducting critical current $I_c$ defining the amplitude of the free energy barrier [32] [46]

$$\Delta F(T) = \frac{\sqrt{6}}{2\pi} \Phi_0 I_c(T) \quad (S2)$$

with the thermal dependence mediated by the critical current $I_c(T) \sim I_c(0)(1 - T/T_c)^{1.5}$ [15].
For a $T < T_c$, so that $\Delta F > k_B T$, the superconducting state is assumed as composed by two kinds of particles, which we have identified as CPs and QPs, so that the total current is

$$I_{in} = I_{qp}(T) + I_s(T) \quad (S3.a)$$

where

$$I_{qp}(T) = I_{in} e^{-\frac{\Delta F(T)}{k_B T}}; \quad I_s = I_{in}\left(1 - e^{-\frac{\Delta F(T)}{k_B T}}\right).$$
(S3.b)

The coefficient $I_{in}$ in eq. (S3.b) is a function of the probability density (eq. (S1)) of a phase slip. An abrupt separation in two kinds of particles does not reflect the realistic situation, where charge carriers pair and depair continuously. It is safe to assume the exponential function as a partition function that sets the density of QPs in relation to the total density, $R(T)/R(T_c) = n_{qp}/(n_{qp} + n_s) = e^{-\frac{\Delta F(T)}{k_B T}}$.

Within the mean-field approach, the screening current dynamics is summarized in Fig.3, where the effect of the external magnetic field consists in the onset of well separated dia- and para-supercurrents in the ideal case. As distinct from that, within the scenario considering the fluctuations of superconductivity due to ongoing pairing and depairing of particles, the dynamics of dia- and para- supercurrents would be disturbed by blurred confinements caused by thermal fluctuations. The degree of smearing of dia- and para- channels could be achieved through the first order fluctuation in $T$ like $\delta n_{s,qp} = (\partial n_{s,qp}/\partial T)\delta T$. Furthermore, from the microscopic point of view, the current should be accounted by considering a distribution of the momenta of both QPs and CPs. These issues are beyond the scope of the present manuscript that aims to provide a simple phenomenologic scenario of interference processes in finite-size mesoscopic rings.